___________________________________________________________

***Deciphering a Sleeping Pathogen: Uncovering Novel Transcriptional Regulators of Hypoxia-Induced Dormancy in Mycobacterium Tuberculosis***

Rohak Jain

Interlake High School

# Abstract

Along the pathogenesis of Mycobacterium Tuberculosis (MTB), hypoxia-induced dormancy is a process involving the oxygen-depleted environment encountered inside the lung granuloma, where bacilli become metabolically inactive and enter into a viable but non-replicating state. Affecting nearly two billion people worldwide, latent TB can linger in the host for indefinite periods of time before resuscitating, which significantly strains the accuracy of treatment options and patient prognosis. Transcription factors thought to mediate this process have only conferred mild growth defects, signaling that our current understanding of the MTB genetic architecture is highly insufficient. In light of these inconsistencies, the objective of this study was to characterize regulatory mechanisms underlying the transition of MTB into dormancy. The project methodology involved a three-part approach: constructing an aggregate hypoxia dataset, inferring a gene regulatory network based on those observations, and leveraging several downstream network analyses to make sense of it all. Results indicated dormancy to be functionally associated with cell redox homeostasis, metal ion cycling, and cell wall metabolism—all of which modulate essential host-pathogen interactions. Additionally, crosstalk between individual regulons (Rv0821c & Rv0144; Rv1152 & Rv2359) was shown to be critical in facilitating bacterial persistence and allowing MTB to gain control over key micronutrients within the cell. Defense antioxidants and nutritional immunity were also identified as avenues to explore further. In providing some of the first insights into the methods utilized by MTB to endure in hypoxia, this research suggests a range of strategies that might aid in improved clinical outcomes of TB treatment.

# Introduction

Ever since its discovery in 1882, Mycobacterium Tuberculosis (MTB) has given rise to a devastating global impact – over 1.5 million lives each year are lost to this lethal pathogen and approximately two billion people, a third of the world's population, are infected with a form of latent TB (Manjelievskaia et al., 2016). Classified as an airborne disease, TB spreads when expelled droplets from the coughs or sneezes of an infected individual are inhaled, prompting foreign bacteria to invade the host immune system, trigger debilitating airway inflammation in the lungs, and spread to other organs in the human body (Russell, 2001). While the high-level pathophysiology of TB infection is widely discussed, an understudied aspect of this process is hypoxia-induced dormancy, where the bacteria persist in a viable, non-replicating state within the host for extended periods of time before returning to their active status (see Figure 2). More specifically, multiple studies have demonstrated that MTB cultures adapt to deoxygenation by entering into a metabolically altered state in the granuloma, a pocket of alveolar macrophages, and waiting until conditions permissive for disease are restored (Rustad et al., 2009). Critically, the formation of dormant cells produces an elevated degree of unpredictability when forecasting the onset of infection, contributes to morphologically distinct lesions in the lungs, and may be responsible for the slow treatment response of patients with active TB. In light of these long-term implications, this study aims to uncover transcriptional agents and novel regulatory patterns responsible for mediating the transition of MTB from hypoxia-induced latency to reaeration.

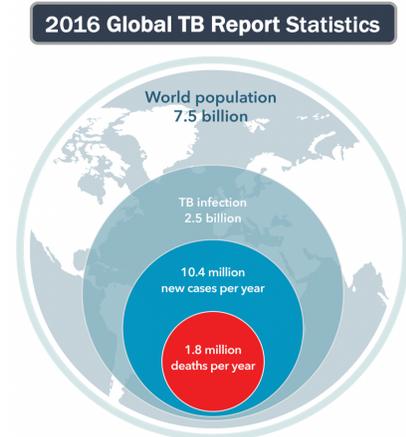

**Figure 1:** Taking a closer look at the scope of TB infection worldwide (Image Credit: Critical Path Institute).

While the high-level objective of this research clearly has a precedent in existing literature, it also addresses critical gaps that have not been properly investigated by similar inquiry-based projects. For instance, the deletion of previously identified transcriptional regulators thought crucial to dormancy (e.g. DosR Regulon) conferred only mild growth defects under hypoxic conditions, indicating that our current understanding of the MTB genetic architecture is highly insufficient (Rustad et al., 2008). Additionally, research in this domain has predominantly relied on error-prone experimental techniques: the extensive usage of electrophoretic mobility shift assays (EMSAs) to locate protein-binding activity could ignore the presence of sample contaminants, while western blotting procedures may fail to account for antibodies interacting with non-intended protein targets (Park et al., 2003; Duque-Correa et al., 2014). Especially in the context of TB, which often resides in parallel with HIV and parasitic infection, such attempts at directed gene disruption and protein localization could give way to questionable accuracy when evaluating results. However, perhaps the biggest barrier when understanding MTB response to hypoxic stress is the incredibly diverse range of modeling techniques that have been used to simulate oxygen depletion. For instance, the Wayne model struggles with reproducibility, the defined hypoxia model involves rapid depletions of oxygen that skip over critical growth periods, and mouse models of necrotic tuberculosis granulomas are difficult to interpret in the context of mammalian disease (Muttucumaru et al., 2004; Harper et al., 2012). As such, computationally shedding light on the network architecture and regulatory mechanisms that underlie the transition from dormancy to reaeration is paramount.

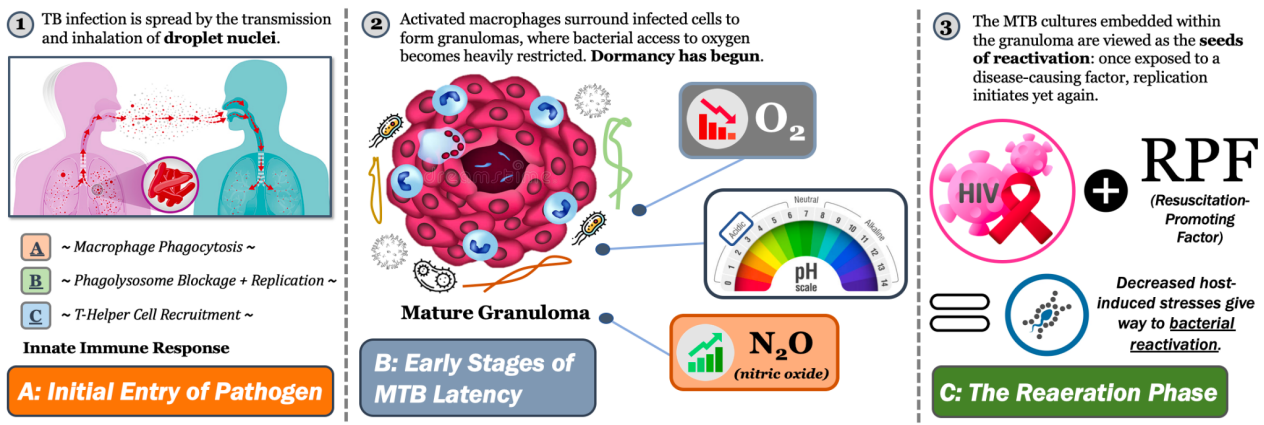

**Figure 2:** A high-level overview of hypoxia-induced dormancy in M. tuberculosis.

Therefore, this study aims to (1) compose an aggregate hypoxia-reaeration dataset from several RNA-Seq and microarray experiments in vivo, (2) infer a gene regulatory network (GRN) based on these observations, and (3) apply a range of downstream analyses to unearth key transcriptional dynamics. By blending experimental platforms with a rigorous systems approach, the underlying interactions that control the transition of MTB in and out of dormancy can be better understood. From characterizing functional states of the pathogen at different treatment stages to investigating how mycobacteria are able to overcome host immune defense, such inquiries can inform the development of improved therapeutics that specifically target phenotypic shifts during disease latency and reactivation.

# Methodology

In terms of the project methodology, a three-phase approach involving data collection, computational modeling, and network analysis was implemented. Dynamic measurements of genome-wide expression over an oxygen gradient were aggregated across multiple sources, which were then filtered through a comprehensive sample similarity and data pre-processing procedure. Subsequently, a set of GRN network inference methods were applied to infer causal relationships among putative transcription factors and gene targets, enabling a holistic look at the regulatory dynamics present across hypoxia and reaeration sampling timepoints. To make sense of the high-dimensional network topology, an assortment of TFOE measures, functional enrichment analyses, and motif detection frameworks were leveraged as appropriate. The flowchart below outlines the general structure of the methodology, and upcoming sections will go into more detail regarding the specifics.

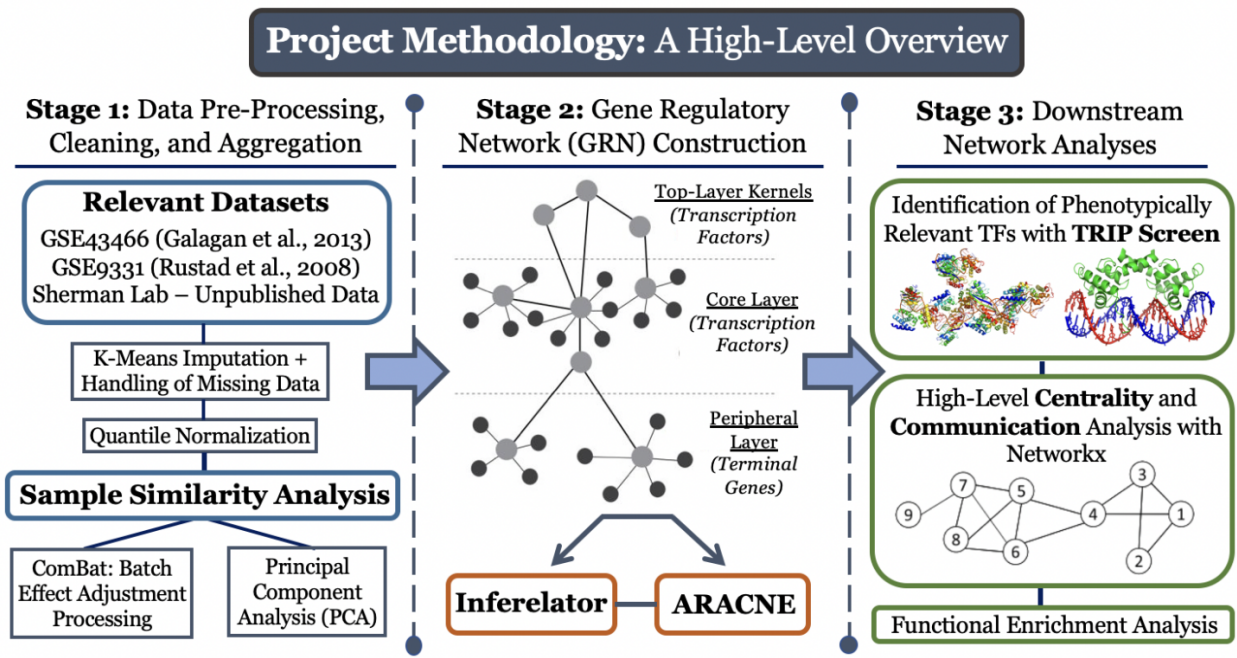

**Figure 3:** A thorough outline of the project methodology.

## Section A: Data Collection and Processing

Throughout this study, three transcriptome datasets that corresponded to hypoxic time course experiments were utilized; two of these were sourced from published papers (GEO Accession Codes: GSE43466 and GSE9331) and the latter was retrieved from an unpublished study conducted by the Sherman Lab at the University of Washington (Rustad et al., 2008; Galagan et al., 2013). To ensure compatibility, special care was taken to assess the experimental designs and cultural conditions used to produce each individual dataset. The following details the general protocol that was found to be consistent across all three investigations:

MTB cultures were grown at 37°C in Middlebrook 7H9 supplemented with ADC and 0.05% Tween (Beckton Dickinson) in rolling culture. Working stocks were expanded from frozen aliquots shortly before experiments began, and the defined hypoxic model was then carried out. Briefly, a 200 ml culture was grown to mid-log phase (A600 = 0.3–0.5), diluted in media to a starting A600 of 0.1, and 500 ml of the diluted culture was transferred to a 1L three-armed spinner flask (Corning). Cultures were constantly stirred at 60 RPM for the duration of the experiment. Low oxygen gas (0.2% $O_2$ with $N_2$ balance) constantly flowed over the culture at 0.15 sq. ft/min. through one arm, while another arm was used as a sampling port. Aliquots were removed at each time point, pelleted at 2000 g for 5 minutes, frozen on dry ice, and stored at −80° until processed for RNA. Critically, the condition timeline involved the deoxygenation treatment in the first seven days (0D to 7D) and the reaeration phase during the period from seven to fourteen days (7D to 14D).

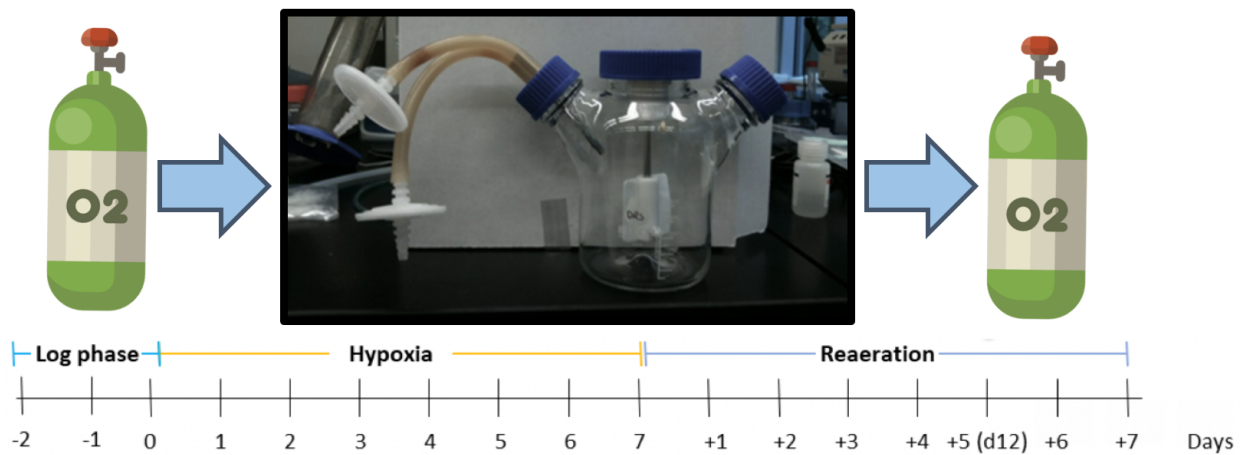

**Figure 4:** Illustrating the generalized experimental setup and protocol (defined hypoxic model) used for data collection across all three studies.

With the raw data in hand, combining it into an aggregated version required establishing commonalities in gene expression patterns such that related timepoints from different studies could be grouped up accordingly. Before getting into all the details, however, quantile normalization was first applied to standardize the statistical properties of each data distribution (Rao et al., 2008). Next, to account for any unwanted technical variability within the experimental samples, the ComBat technique was leveraged to perform batch effect adjustment, primarily because its non-parametric Empirical Bayes frameworks have been shown to work well on microarray datasets (Johnson et al., 2007). KNN-based imputation was also implemented to fill in any null or missing values.

The cleaned dataframe was then inputted into a Principal Component Analysis (PCA) pipeline, which aimed to rigorously test for likeness among related timepoints (Abdi & Williams, 2010). Within the combined dataset, a PCA Plot (PC1 vs. PC2) and a Biplot (PCs + Loading Scores) were both generated as a means of cleanly illustrating the spatial similarity between the sample clusters. Additionally, it's worth noting that numerous timepoints representative of distinct stages along the timeline of hypoxia (ranging from 12HR to 7D post-deoxygenation) were used. Finally, Euclidean and Manhattan point-distance measures, along with a dendrogram visualization, further illustrated the high-level relationships between the various samples. For reference, all analysis was performed in Python with ScikitLearn's 'PCA' functionality, Bioinfokit's clustering visualization tools, and inbuilt Pandas utilities.

As evidenced by the small acute angles shared between the biplot vectors, the negligible difference in PC1 contribution among the samples (x-axis tick spacings of 0.02), and the tightly-bound clustering relationships displayed in the dendrogram, the variables in the PCA analysis proved to be highly correlated. Perhaps more interestingly, these similarities were seemingly coordinated with the time series, with the early (12HR, 1D), intermediate (2D, 3D, 4D), and late (5D, 7D) samples being grouped alongside each other. As such, the conditions for data aggregation were satisfied and measurements across all three studies were combined accordingly. For more information on the PCA analysis, please reference the results delineated in Figure 5.

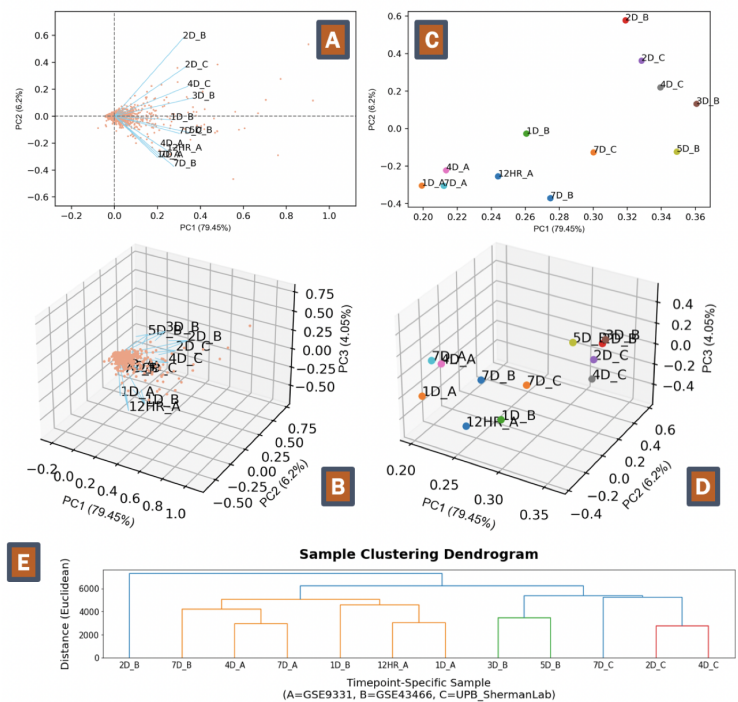

**Figure 5:** Results from the PCA Analysis, with the 2D- and 3D-versions of the Biplot and PCA Plot (along with the clustering dendrogram) shown clearly. Please note that the letter-based naming scheme corresponds to study-specific samples (A = 'GSE9331,' B = 'GSE43466,' C = 'UPB_ShermanLab').

## Section B: Gene Regulatory Network Design and Implementation

Thanks to the availability of high-throughput gene expression data in recent times, gene regulatory networks find themselves at the forefront of cutting-edge biomedical research. Allowing for the characterization of transcriptional relationships, gene-gene interactions, and other aspects of diverse biological systems, GRNs can be applied in several contexts (Emmert-Streib et al., 2014). With reference to TB, there clearly exists a precedent for this approach, with high-resolution network models being used to understand response to growth arrest, global gene regulation, and viral persistence (Balázsi et al., 2008; Peterson et al., 2014; Wang et al., 2011). Interestingly enough, Galagan and colleagues built upon existing ChIP-Seq data to inform the development of a draft GRN; ultimately, this analysis revealed aspects of the hypoxic response and highlighted alterations in lipid metabolism as a result of oxygen availability (Galagan et al., 2013). However, the project only featured fifty transcription factors and was not greatly validated by related studies.

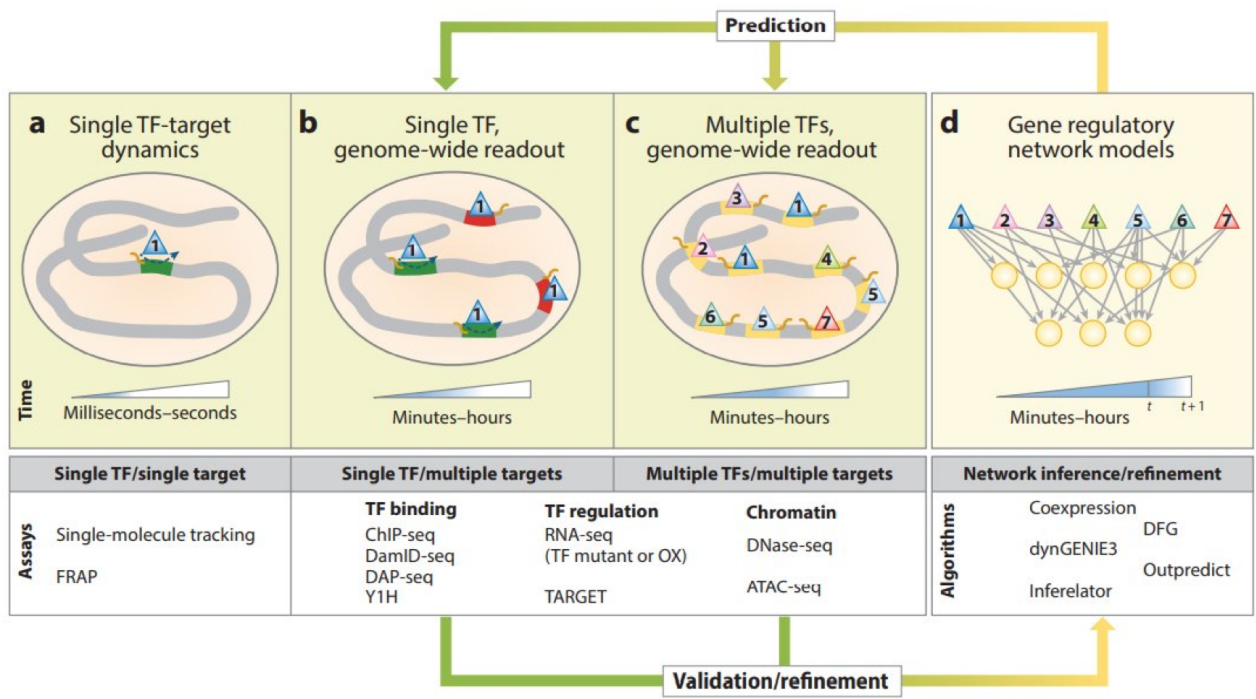

**Figure 6:** An overview of the gene regulatory network (GRN) pipeline. (Image Credit: Alvarez, JM. (2021) *Time-Based Systems Biology Approaches to Capture and Model Dynamic Gene Regulatory Networks* [publication]).

To this end, two GRN construction approaches were utilized, since structural components across both networks could be synthesized to determine the strength of TF-gene interactions. For reference, the details of each method can be found below:

1. **Inferelator:** Logistically speaking, the Inferelator 3.0 pipeline computes TF activity with a prior knowledge network and, based on the extracted results, fits a regularized regression model on input scRNAseq data to learn new regulatory edges (Skok Gibbs et al., 2022) . Additionally, Inferelator builds off of gold standard files with experimentally determined TF-target interactions, which results in improved prediction accuracy. As such, this tool has been found to work exceptionally well on datasets containing millions

**6**

of cell samples, provides scalability, and retains the ability to examine bulk gene expression values.

2. **ARACNE:** On a high-level, ARACNE identifies candidate interactions by estimating the mutual information (MI) between gene expression profiles and filters the resultant MIs using a threshold derived from the null-hypothesis of two independent genes (Margolin et al., 2006). Subsequently, a property termed the data processing inequality (DPI) is used to remove the majority of indirect interactions / false positives contained in the network. As a result, the low computational complexity and assumption-free nature of ARACNE allows for the recovery of transcriptional relationships with high confidence.

After the GRNs had been successfully built, the ranking dataframe (strength of TF-gene edges) and the complete connectivity matrix were saved to external variables that could be accessed later on. A total of 3889 genes, along with 207 annotated transcription factors, were represented in the final networks.

### Section C: Downstream Network Analyses

High-level network visualizations were carried out with the NetworkX and RedeR software, where major communication hubs and regulatory circuits could be easily mapped out (Castro et al., 2013). Betweenness centrality, along with in- and out-degree counts, was used to quantify the degree of connectedness for any given transcription factor. As a sanity check, TF-gene interactions were cross-checked against experimental transcription factor overexpression (TFOE) data taken from mycobacterial broth cultures in growth media.

In addition, network motifs were inferred from the raw GRN data with the Pymfinder package, which uses a modified version of the 'mfinder' engine to account for interaction strengths and outline a set of motif role and participation data (Mora et al., 2018). To characterize the functional role of genes and TFs that had been heavily implicated in the network, enrichment analysis was performed with Mycobrowser gene annotation data and the GSEA Python package (Kapopoulou et al., 2011; Fang et al., 2022).

# Results

### *MTB **Hypoxia GRN** Reveals Interesting TF Dynamics*

Utilizing the aggregate dataset, the dual-method (ARACNE & Inferelator) GRN inference approach was applied, revealing an intricate network of TF-gene relations. Functional enrichment analysis with GSEAPy's Enrichr plugin suggested that the network components were most directly associated with zinc ion metabolism, the dTDP biosynthetic process, defense response to bacterium, cell redox homeostasis, and deoxyribonuclease activity. It is interesting to see how several of these linkages, particularly in relation to essential micronutrients and cellular maintenance, have previously been shown to mediate MTB adaptations within the granuloma. For instance, ensuring genomic stability – a goal often achieved through imbalanced toxin-antitoxin (TA) systems, slowed energy production, and the manipulation of cytokine responses – is critical in prolonging the persistence of infection (Peddireddy et al., 2017). Moving past the high-level functional groupings expressed within the GRNs, gleaning further insight into specific TF-TF and TF-gene relationships was another important project objective. For this purpose, Figure 7 sheds light on the consensus network topology and spotlights dense regulatory hubs with degree-rank and centrality plots.

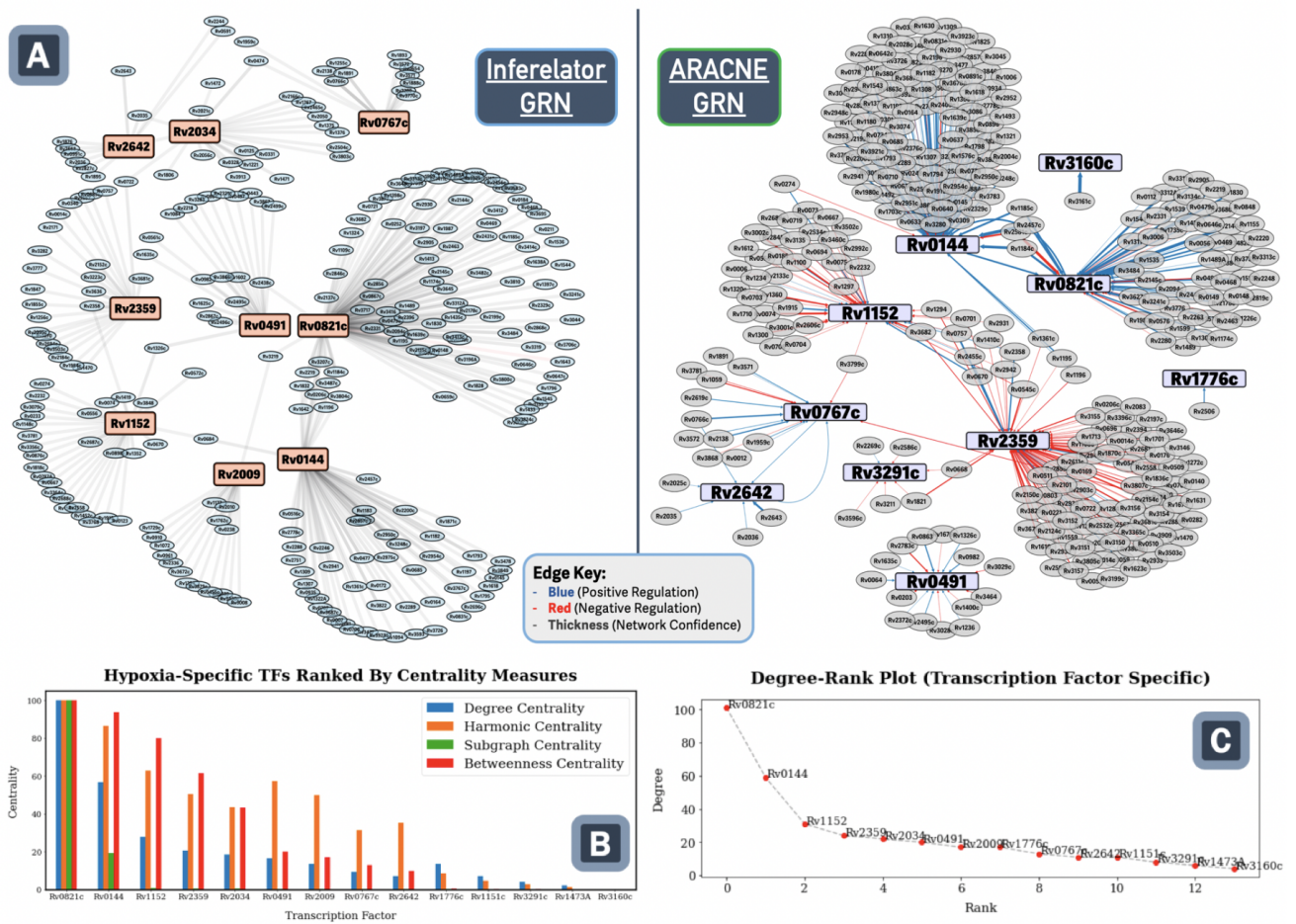

**Figure 7:** A graphic of the hypoxia GRNs and the most influential TFs ranked in terms of their centrality measure and in/out degree connections. Note that the regulators shown here stem directly from the TRIP-identified growth abundance and defect TFs under hypoxic conditions.

## *Investigating the **Rv0144-Rv0821c Crosstalk***

One compelling TF association involved Rv0144 and Rv0821c, seeing that they ranked highest in terms of centrality across the GRNs and individually formed extensive communities of target genes and proteins. Interestingly enough, the inactivation of Rv0821c (PhoY2) was shown to cause a considerable defect in the mycobacterial persistence phenotype and lead to increased antibiotic resistance (Shi & Yang, 2010). Aside from that, PhoY2 has also been implicated in intracellular inorganic phosphate homeostasis and balanced energy-redox state, both of which are critical in the metabolic shift of MTB under stress conditions (Wang et al., 2013; Namugenyi et al., 2017). In regards to phosphate depletion, MTB must frequently contend with a phosphate-limited environment in macrophage phagosomes and, as such, PhoY2's role in controlling the phosphate starvation response can restrict attempts to stall bacterial growth in a dose-dependent manner (Rifat et al., 2009). Similarly, Rv0144 has previously been demonstrated to be regulated by RelA, known to be essential in establishing persistent infection in mice (Dahl et al., 2003). As such, the strong ties between Rv0144 and Rv0821c could serve as a dual mechanism of persistence that contributes to successful latent infection.

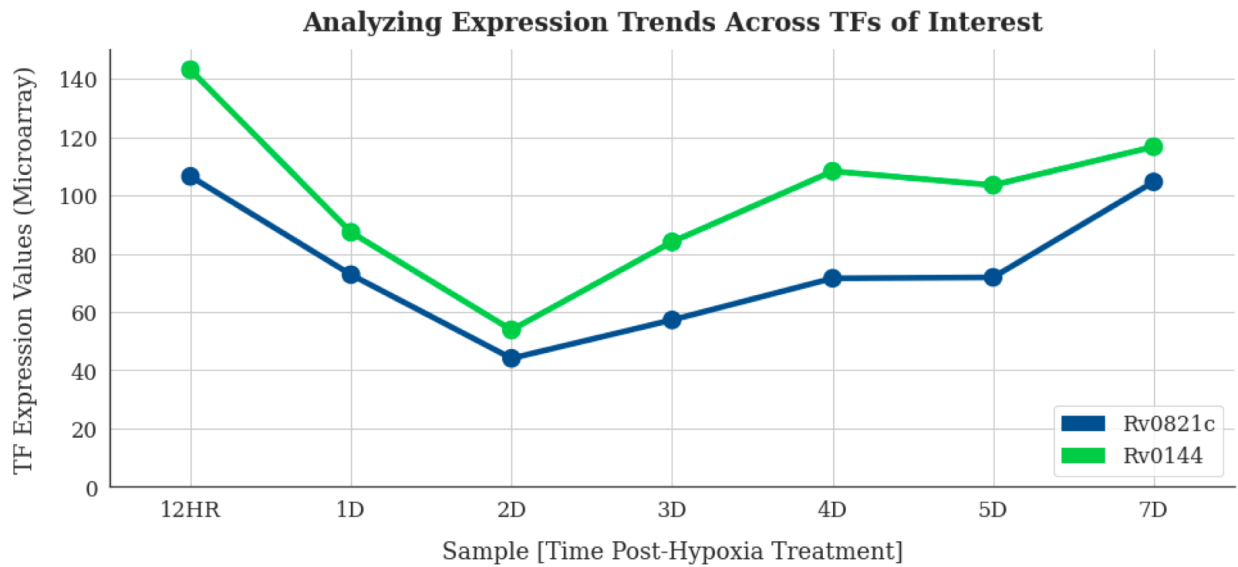

**Figure 8:** A Rv0821-Rv0144 Expression Plot.

Revisiting prior analyses, the integrity of this relationship was substantially strengthened. In a comparison of the raw trends in gene expression across both TFs, a clear correlation was found with a Pearson's Correlation Coefficient of 0.915. Although the general expression pattern saw an initial drop-off in the early stages post-hypoxia, it quickly recovered in the later timepoints and returned to elevated levels. Additionally, when the shared target components of Rv0821c and Rv0144 were evaluated, they appeared to be significantly correlated with cell wall structure and repair (see Table 2). Since the lipid- and carbohydrate-rich layers of the MTB cell wall allow for the creation of a selective permeability barrier, in which hydrophilic compounds are protected against, the regulation of cell wall biosynthesis enzymes could stimulate pathogenicity (Brennan, 2003). With this in mind, the Rv0144-Rv0821c system's immediate downregulation and subsequent return to steady-state might suggest a fine-tuned balancing act of sorts, where intracellular membrane transport and cell wall enzymatic activity are preserved despite being originally attenuated. Such a mechanism could bypass immune defense and pave the way for the proliferation of virulence within the host.

**Shared Target Components** Between Rv0821c and Rv0144

| Target | Functional Description | Category |
|---|---|---|
| Rv1184c | Essential for PAT lipid biosynthesis, which is a significant constituent of the **mycobacterial cell wall**. | *Cell Wall and Cell Processes* |
| Rv0206c | MmpL3 protein is a transmembrane transporter of mycolic acid; long chain fatty acids found in the **lipid-rich cell walls** of tuberculosis bacterium. | *Cell Wall and Cell Processes* |
| Rv3804c | Refers to proteins of the antigen 85 complex that contribute to the biogenesis of trehalose dimycolate, a dominant structure required for **cell wall integrity**. | *Lipid Metabolism* |
| Rv3487c | Lipolytic enzyme LipF involved in cellular metabolism. | *Intermediary Metabolism and Respiration* |
| Rv2219 | Probable conserved transmembrane protein. | *Cell Wall and Cell Processes* |
| Rv1832 | Glycine cleavage system that catalyzes the degradation of glycine, which has been implicated in the biosynthesis of peptidoglycan and other **cell wall structural components**. | *Intermediary Metabolism and Respiration* |
| Rv1196 | Resembles PPE18, a **cell wall associated protein** that is involved in inflammatory response and cytokine manipulation. | *PE/PPE* |

**Table 2:** Examining discernable trends in the functional roles of shared target components between Rv0821c and Rv0144; gene annotation performed with the Enrichr API and GSEAPy.

## Characterizing the *Rv2359-Rv1152 Relationship*

Another intriguing regulatory circuit consisted of Rv2359, a probable zinc uptake protein Zur, and Rv1152, a transcriptional protein implicated in vancomycin antibiotic resistance (Zeng et al., 2016). Combined with a strong expression correlation (Pearson's Correlation Coefficient = 84.6), both regulators also held an exceedingly similar TRIP phenotype, with a roughly four-fold downregulation under hypoxia stress.

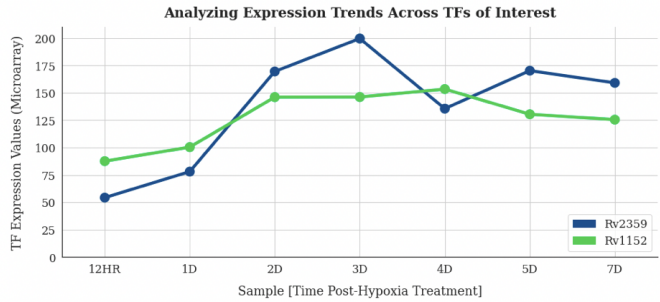

**Figure 9:** Illustrating the consistent alignment between Rv2359 and Rv1152 in both expression- and phenotype-based data metrics.

Furthermore, it is worth noting that Rv2359, along with a cofactor Rv2358, have been shown to collectively form a zinc homeostasis operon whose role is to remove metal ions at high, toxic concentrations (Milano et al., 2004). Under the proposed model, low to moderate zinc levels result in autorepression of the operon by Rv2359 and a failure to initiate transcription; elevated zinc levels, however, cause metal binding by the sensor protein Rv2358, which lowers its DNA binding affinity and enables RNA Polymerase to load and begin transcription of downstream effector genes (Canneva et al., 2005). Ultimately, then, the cotranscription of Rv2358 and Rv2359 likely reflects a necessary element involved in compensating between the import and export of zinc. Incidentally, in both of the hypoxia GRNs, Rv2358 was found to only share edges with Rv2359 and Rv1152 among the set of candidate hypoxia regulators.

In the context of hypoxia, prolonged zinc limitation can result in anticipatory pathogenic adaptations against impending immune attack. Moreover, zinc sequestration may override the process of nutritional immunity, in which the host immune system limits the production of micronutrients to control pathogen growth (Dow et al., 2021). Considering that Rv1152 alters the cell wall permeability of MTB to acid and surface stress, its strong ties to Rv2358 and Rv2359 could suggest a membrane transport mechanism used to mediate zinc concentrations within the hypoxic granuloma. The suppression of Rv1152 and Rv2359 in hypoxia reinforces this hypothesis, as a lack of zinc control factors would lead to detoxification of pro-inflammatory reactive oxygen species (ROS), changes in lipodome composition, and other protective measures (Voskuil et al., 2011). Similar functional groupings can be seen in the shared network genes between Rv1152 and Rv2359, which were connected to ribosomal RNA-binding, DNA repair, intracellular transport, cell wall processes, and a range of biosynthetic pathways. In agreement with these observations, the Rv1152-Rv2359 dynamic could not only afford MTB the ability to adapt to diverse environments within the host, but also might constitute an anticipatory response signaled by low $Zn^{2+}$ in preparation for imminent phagocytosis and transit through the hypoxic caseum.

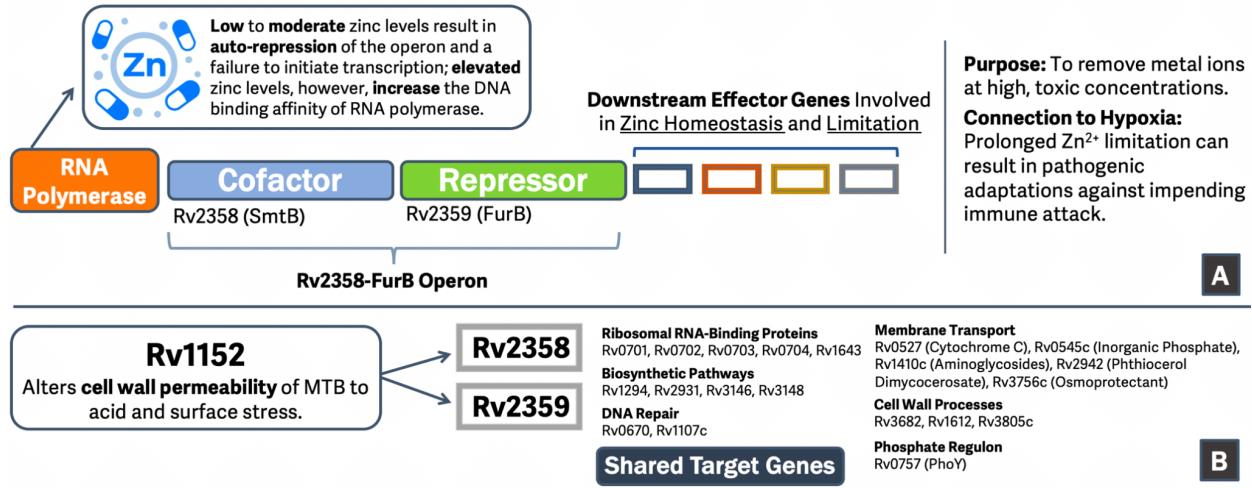

**Figure 10:** A graphic of the multi-faceted linkages between Rv2359 and Rv1152.

# Discussion + Concluding Thoughts

In this report, a data-driven approach toward studying latent TB infection revealed insights that comment on the elaborate adaptive mechanisms of bacteria when faced with oxygen depletion. In contrast to previous work in this area, which has predominantly relied on low-resolution hypoxia gene expression profiling, an aggregate dataset spanning three individual studies was leveraged to significantly augment the accuracy of subsequent analyses. Following from the data collection protocol, a two-part GRN construction process (Inferelator & ARACNE) indicated a set of high-level functional categories associated with the hypoxia gene pool (see Table 3). These annotations included redox homeostasis, nutrient transport across the cell membrane, and other stress response protein families that likely facilitate the extension of virulence into the nonreplicating state. Further inquiry into the network called attention to the interplay between Rv0144 and Rv0821c, eventually unveiling a channel of persistence brought about by the proliferation of cell wall synthesis components. In addition, the crosstalk among Rv1152 and Rv2359 reflected a metal ion-respondent homeostasis mechanism implicated in guiding the anticipative adaptations of MTB to incoming immune attack and mitigating the effects of oxidative stressors (e.g. TNF-Alpha).

| GO Term | Overlap | P-Value | Genes | Phenotypic Relevance |
|---|---|---|---|---|
| Peptidoglycan Biosynthetic Process | 8/15 | 0.003567 | Rv2154c; Rv1086; Rv3682; Rv3794; Rv2152c; Rv0483; Rv0050; Rv1018c. | The peptidoglycan layer is essential for maintaining cellular integrity and forming a permeability barrier. |
| Proton-Transporting ATP Synthase Activity | 6/8 | 0.034982 | Rv1309; Rv1311; Rv1307; Rv1310; Rv1308; Rv1306. | Protonmotive force is required for maintaining ATP homeostasis and viability of hypoxic MTB. |
| Cell Redox Homeostasis | 5/12 | 0.002969 | Rv1470; Rv1471; Rv0688; Rv1324; Rv1677. | Preservation of an appropriate redox balance is critical to the persistence of MTB. |
| Fatty Acid Biosynthetic Process | 7/17 | 0.048612 | Rv3825c; Rv1484; Rv2524c; Rv0533c; Rv1094; Rv2244; Rv2246. | Macrophage fatty acid metabolism is needed to supplement MTB survival in hypoxia. |
| Response to Stress | 8/14 | 0.013853 | Rv3223c; Rv2028c; Rv3134c; Rv2374c; Rv2624c; Rv0576; Rv0982; Rv2035. | An indicator that bacteria are sensing and adapting to the anaerobic environment. |

*\* Enrichment analysis was performed with the Enrichr API of GSEAPy; an adjusted P-Value cutoff of <= 0.05 was used to determine statistical significance.*

**Table 3:** Interesting results from the Gene Ontology explorations.

However, despite the intriguing findings at play, it's also worth noting some of the key limitations that hinder the strength of any derived conclusions and hypotheses. For instance, the dataset was solely based on the defined hypoxic model, which involves the exposure of bacilli to a consistent low oxygen tension (1-2%) over a predetermined time period. Still, though, questions remain as to how well these in vitro conditions account for the subtle changes in molecular signature and metabolism experienced by hypoxic bacteria. As such, sampling during critical transition periods could be substantially weakened, effectively preventing the real-time monitoring of the MTB transcriptional state across an accurate $O_2$ gradient. Another shortcoming concerns the lack of gold standard data, seeing as the recovery of real-world scRNAseq data enumerating known TF-target gene interactions has been shown to far surpass the performance of prior knowledge GRNs (Banf & Rhee, 2017). Without supplemental information, the prediction of TF activity enters into a nebulous gray area and the likelihood of redundancy in the final network is raised considerably.

Going forward, future directions could be filtered down multiple avenues. To begin with, reaeration data measuring gene expression in the five-day interval post-hypoxia treatment (7D to 12D) could be incorporated to catalog other MTB physiological adjustments during reintroduction to the stationary phase. Moreover, the DREM 2.0 GRN inference approach is another target of interest, as it identifies bifurcation points that follow transitions between coordinated regulatory programs and gene states (Schulz et al., 2012). Putting both of these refinements in concert with one another will allow for a more holistic approach covering a wider range of latency stages and growth periods.

The vast reservoir of latent TB, encompassing nearly 1.8 billion people, continues to be a source of reactivation disease. Accordingly, pinpointing novel regulatory machinery with characteristic features is essential in heightening the accuracy of patient prognosis. In providing some of the first insights into the mechanisms utilized by MTB to endure in a quiescent state, this study suggests a variety of predictive and rational strategies that might aid in improved clinical outcomes of TB treatment. Although follow-up experiments will be needed to validate computationally-identified trends, this project takes a fundamental step in the ongoing fight against TB – a milestone that might be more valuable now than ever before.

# References

Abdi, H., & Williams, L. J. (2010). Principal component analysis. *Wiley interdisciplinary reviews: computational statistics*, *2*(4), 433-459.

Balázsi, G., Heath, A. P., Shi, L., & Gennaro, M. L. (2008). The temporal response of the Mycobacterium tuberculosis gene regulatory network during growth arrest. *Molecular systems biology*, *4*(1), 225.

Banf, M., & Rhee, S. Y. (2017). Computational inference of gene regulatory networks: approaches, limitations and opportunities. *Biochimica et Biophysica Acta (BBA)-Gene Regulatory Mechanisms*, *1860*(1), 41-52.

Brennan, P. J. (2003). Structure, function, and biogenesis of the cell wall of Mycobacterium tuberculosis. *Tuberculosis*, *83*(1-3), 91-97.

Canneva, F., Branzoni, M., Riccardi, G., Provvedi, R., & Milano, A. (2005). Rv2358 and FurB: two transcriptional regulators from Mycobacterium tuberculosis which respond to zinc. *Journal of bacteriology*, *187*(16), 5837-5840.

Castro, M. A., Wang, X., Fletcher, M. N., Markowetz, F., & Meyer, K. B. (2013). Vignette for RTN: reconstruction of transcriptional networks and analysis of master regulators.

Dahl, J. L., Kraus, C. N., Boshoff, H. I., Doan, B., Foley, K., Avarbock, D., ... & Barry III, C. E. (2003). The role of RelMtb-mediated adaptation to stationary phase in long-term persistence of Mycobacterium tuberculosis in mice. *Proceedings of the National Academy of Sciences*, *100*(17), 10026-10031.

Dow, A., Sule, P., O'Donnell, T. J., Burger, A., Mattila, J. T., Antonio, B., ... & Prisic, S. (2021). Zinc limitation triggers anticipatory adaptations in Mycobacterium tuberculosis. *PLoS pathogens*, *17*(5), e1009570.

Duque-Correa, M. A., Kühl, A. A., Rodriguez, P. C., Zedler, U., Schommer-Leitner, S., Rao, M., ... & Reece, S. T. (2014). Macrophage arginase-1 controls bacterial growth and pathology in hypoxic tuberculosis granulomas. *Proceedings of the National Academy of Sciences*, *111*(38), E4024-E4032.

Emmert-Streib, F., Dehmer, M., & Haibe-Kains, B. (2014). Gene regulatory networks and their applications: understanding biological and medical problems in terms of networks. *Frontiers in cell and developmental biology*, *2*, 38.

Fang, Z., Liu, X., & Peltz, G. (2022). GSEApy: a comprehensive package for performing gene set enrichment analysis in Python. *Bioinformatics*.

Galagan, J. E., Minch, K., Peterson, M., Lyubetskaya, A., Azizi, E., Sweet, L., ... & Schoolnik, G. K. (2013). The Mycobacterium tuberculosis regulatory network and hypoxia. *Nature*, *499*(7457), 178-183.

Harper, J., Skerry, C., Davis, S. L., Tasneen, R., Weir, M., Kramnik, I., ... & Jain, S. K. (2012). Mouse model of necrotic tuberculosis granulomas develops hypoxic lesions. *Journal of Infectious Diseases*, *205*(4), 595-602.

Johnson, W. E., Li, C., & Rabinovic, A. (2007). Adjusting batch effects in microarray expression data using empirical Bayes methods. *Biostatistics*, *8*(1), 118-127.

Kapopoulou, A., Lew, J. M., & Cole, S. T. (2011). The MycoBrowser portal: a comprehensive and manually annotated resource for mycobacterial genomes. *Tuberculosis*, *91*(1), 8-13.

Ma, S., Morrison, R., Hobbs, S. J., Soni, V., Farrow-Johnson, J., Frando, A., ... & Sherman, D. R. (2021). Transcriptional regulator-induced phenotype screen reveals drug potentiators in Mycobacterium tuberculosis. *Nature microbiology*, *6*(1), 44-50.

Manjelievskaia, J., Erck, D., Piracha, S., & Schrager, L. (2016). Drug-resistant TB: deadly, costly and in need of a vaccine. *Transactions of the Royal Society of Tropical Medicine and Hygiene*, *110*(3), 186-191.

Margolin, A. A., Nemenman, I., Basso, K., Wiggins, C., Stolovitzky, G., Favera, R. D., & Califano, A. (2006, March). ARACNE: an algorithm for the reconstruction of gene regulatory networks in a mammalian cellular context. In *BMC bioinformatics* (Vol. 7, No. 1, pp. 1-15). BioMed Central.

Milano, A., Branzoni, M., Canneva, F., Profumo, A., & Riccardi, G. (2004). The mycobacterium tuberculosis rv2358–furB operon is induced by zinc. *Research in microbiology*, *155*(3), 192-200.

Mora, B. B., Cirtwill, A. R., & Stouffer, D. B. (2018). pymfinder: a tool for the motif analysis of binary and quantitative complex networks. *BioRxiv*, 364703.

Muttucumaru, D. N., Roberts, G., Hinds, J., Stabler, R. A., & Parish, T. (2004). Gene expression profile of Mycobacterium tuberculosis in a non-replicating state. *Tuberculosis*, *84*(3-4), 239-246.

Namugenyi, S. B., Aagesen, A. M., Elliott, S. R., & Tischler, A. D. (2017). Mycobacterium tuberculosis PhoY proteins promote persister formation by mediating Pst/SenX3-RegX3 phosphate sensing. *MBio*, *8*(4), e00494-17.

Park, H. D., Guinn, K. M., Harrell, M. I., Liao, R., Voskuil, M. I., Tompa, M., ... & Sherman, D. R. (2003). Rv3133c/dosR is a transcription factor that mediates the hypoxic response of Mycobacterium tuberculosis. *Molecular microbiology*, *48*(3), 833-843.

Peddireddy, V., Doddam, S. N., & Ahmed, N. (2017). Mycobacterial dormancy systems and host responses in tuberculosis. *Frontiers in immunology*, *8*, 84.

Peterson, E. J., Reiss, D. J., Turkarslan, S., Minch, K. J., Rustad, T., Plaisier, C. L., ... & Baliga, N. S. (2014). A high-resolution network model for global gene regulation in Mycobacterium tuberculosis. *Nucleic acids research*, *42*(18), 11291-11303.

Rao, Y., Lee, Y., Jarjoura, D., Ruppert, A. S., Liu, C. G., Hsu, J. C., & Hagan, J. P. (2008). A comparison of normalization techniques for microRNA microarray data. *Statistical applications in genetics and molecular biology*, *7*(1).

Rifat, D., Bishai, W. R., & Karakousis, P. C. (2009). Phosphate depletion: a novel trigger for Mycobacterium tuberculosis persistence. *The Journal of infectious diseases*, *200*(7), 1126-1135.

Russell, D. G. (2001). Mycobacterium tuberculosis: here today, and here tomorrow. *Nature reviews Molecular cell biology*, *2*(8), 569-578.

Rustad, T. R., Harrell, M. I., Liao, R., & Sherman, D. R. (2008). The enduring hypoxic response of Mycobacterium tuberculosis. *PloS one*, *3*(1), e1502.

Rustad, T. R., Sherrid, A. M., Minch, K. J., & Sherman, D. R. (2009). Hypoxia: a window into Mycobacterium tuberculosis latency. *Cellular microbiology*, *11*(8), 1151-1159.

Schulz, M. H., Devanny, W. E., Gitter, A., Zhong, S., Ernst, J., & Bar-Joseph, Z. (2012). DREM 2.0: Improved reconstruction of dynamic regulatory networks from time-series expression data. *BMC systems biology*, *6*(1), 1-9.

Shi, W., & Zhang, Y. (2010). PhoY2 but not PhoY1 is the PhoU homologue involved in persisters in Mycobacterium tuberculosis. *Journal of Antimicrobial Chemotherapy*, *65*(6), 1237-1242.

Skok Gibbs, C., Jackson, C. A., Saldi, G. A., Tjärnberg, A., Shah, A., Watters, A., ... & Bonneau, R. (2022). High-performance single-cell gene regulatory network inference at scale: the Inferelator 3.0. *Bioinformatics*, *38*(9), 2519-2528.

Voskuil, M. I., Bartek, I. L., Visconti, K., & Schoolnik, G. K. (2011). The response of Mycobacterium tuberculosis to reactive oxygen and nitrogen species. *Frontiers in microbiology*, *2*, 105.

Wang, C., Mao, Y., Yu, J., Zhu, L., Li, M., Wang, D., ... & Gao, Q. (2013). PhoY2 of mycobacteria is required for metabolic homeostasis and stress response. *Journal of bacteriology*, *195*(2), 243-252.

Wang, X., Wang, H., & Xie, J. (2011). Genes and regulatory networks involved in persistence of Mycobacterium tuberculosis. *Science China Life Sciences*, *54*(4), 300-310.

Zeng, J., Deng, W., Yang, W., Luo, H., Duan, X., Xie, L., ... & Xie, J. (2016). Mycobacterium tuberculosis Rv1152 is a novel GntR family transcriptional regulator involved in intrinsic vancomycin resistance and is a potential vancomycin adjuvant target. *Scientific reports*, *6*(1), 1-12.

NOTE: All images, unless specified otherwise, were produced by the student researcher.